 \documentclass{cpp2e}

\newcommand{\be}{\begin{equation}}
\newcommand{\ee}{\end{equation}}
\newcommand{\bea}{\begin{eqnarray}}
\newcommand{\eea}{\end{eqnarray}}

 \shorttitle{Plasma Reflectivity} 
 \shortauthor{H. Reinholz} 
 \contvolum{vol} 
 \contyear{year} 
 \contnumber{num} 
 \contpages{p$-$p} 
 \title{ Reflectivity of Shock Compressed Xenon Plasma} 
 \author{ H. Reinholz$^{\rm a,b}$, G. R\"opke$^{\rm b}$, 
 A. Wierling$^{\rm b}$,\\
  V. Mintsev$^{\rm c}$ , and V. Gryaznov$^{\rm c}$ }
 \address{
 $^{\rm a}$ University of Western Australia, Dept. of Physics,
 35 Stirling Highway, Crawley, WA 6009, Australia\\
 $^{\rm b}$ Universit\"at Rostock, FB Physik, D-18051
   Rostock, Germany\\
 $^{\rm c}$ Institute of Problems of Chemical Physics,
   Chernogolovka, Moscow Reg., 142432 Russia\\
 e-mail:heidi@physics.uwa.edu.au}

\begin{document}

\setcounter{page}{1}
 \makeheadings
 \maketitle

\begin{abstract}
  Experimental results \cite{Mint} 
  for the reflection coefficient of shock-compressed
  dense Xenon plasmas at pressures of 1.6 -- 17 GPa and temperatures 
  around 30\,000 K using a
  laser beam with $\lambda = 1.06\,\, \mu$m   are compared with
  calculations based on different theoretical approaches to the
  dynamical collision frequency. It is found that a reasonable
  description can be given assuming a spatial electron density
  profile corresponding to a finite width of the shock wave front of
  about $2\cdot10^{-6}$ m.
\end{abstract}

\section{Reflectivity measurements}

Experiments with an explosively driven generator of 
shock waves to produce dense nonideal Xenon plasmas were reported  
in \cite{Mint}. The reflectivity was measured using a laser system
with the wave length $\lambda_{l} = 1.06\,\, \mu$m. The results of the
experiments are shown in Tab.\,1. The thermodynamic parameters of the
plasma were determined from the measured shock wave velocity. 
The plasma composition was calculated within a chemical picture 
\cite{ebel}. Working with a grand canonical ensemble \cite{lik} 
virial corrections have been taken into account due to charge-charge 
interactions (Debye approximation). Short range repulsion 
of atoms and ions (including multiply charged positive ions) was 
considered via second and third virial coefficients within a 
virial expansions. In the parameter range of the shock wave experiments, 
derivations of up to 20 \% for the 
composition have been 
obtained depending on the approximations for the equation of state.
 This is within the accuracy of the 
experimental values of the reflectivity. \\

\begin{table}
\begin{center}
\begin{tabular}{|r|l|c|l|l|l|}
\hline
$P$/ GPa & $T$/ K &   $\rho$/ g\,cm$^{-3}$ & $n_{e}$/ cm$^{-3}$ &
$n_{a}$/cm$^{-3}$ & $R$ 
\\ \hline
1.6  &   30050 &  0.51 &   1.8$\times 10^{21}$ & 6.1$\times 10^{20}$ & 0.096\\
3.1  &   29570 &  0.97 &   3.2$\times 10^{21}$& 1.4$\times 10^{21}$ & 0.12\\
5.1  &   30260 &  1.46 &   4.5$\times 10^{21}$ & 2.2$\times 10^{21}$ & 0.18\\
7.3  &   29810 &  1.98 &   5.7$\times 10^{21}$ & 3.5$\times 10^{21}$ & 0.26\\
10.5 &   29250 &  2.70 &   7.1$\times 10^{21}$ & 5.4$\times 10^{21}$ & 0.36\\
16.7 &   28810 &  3.84 &   9.1$\times 10^{21}$ & 8.6$\times 10^{21}$ & 0.47\\
\hline
        \end{tabular} \end{center}
        \caption{Experimental results for the reflectivity $R$ of
          Xenon plasmas at given parameter values: pressure $P$,
          temperature $T$, mass density $\rho$, free electron number
          density $n_e$ and density of neutral atoms $n_a$.}
        \label{tab1}
\end{table}

It was suggested that the measurement of reflectivity allows to
determine the concentration of free electrons in the plasma. However,
investigating these data \cite{Mint}, it was not possible to find a 
direct relationship 
between the values of free electron  density $n_e$ and the 
reflectivity $R$. The reflection coefficient increases smoothly with the free
electron density. It approaches only slowly values characteristic for 
metals, 
although the critical density for metallic behaviour  
$n_e^{\rm cr}=1.02\times 10^{21} {\rm cm}^{-3}$, where the plasma
frequency, $\omega_{\rm pl}=\sqrt{n_{e}e^{2}/(\epsilon_{0}m_{e})}$ 
with $m_{e}$ the electron mass,
coincides with the frequency $\omega_{l}$ of the probing laser pulse, is
exceeded even at the lowest density. 
However, this critical density is taken from the RPA approximation for 
the dielectric function in the long--wavelength limit
$\epsilon\,=\,1-\omega_{\rm pl}^{2}/\omega_{l}^{2}$ where total 
reflection occurs due to a vanishing dielectric function. Taking 
collisions into account this will be smoothed out. On the other hand, 
a sharp boundary between plasma and undisturbed gas in front of 
shock wave is assumed.

In \cite{Mint}, the spatial structure of the ionizing shock wave was
discussed, showing three characteristic zones, which may influence the
electromagnetic wave propagation. In a precursor zone, the gas is
heated, but the influence on the wave propagation is small. In the
region of the shock wave front a steep increase of the free electron
density is expected. The width of the wave front is determined by
relaxation processes in the plasma.  It was estimated to
be of the order $d \approx 10^{-7}$ m \cite{Mint}, which is one order 
of magnitude less than the
laser wave length.  Under this condition $d/\lambda \ll 1$, the laser
beam reflection was assumed to be determined by the electron properties
of the plasma behind the shock wave front. An expression derived from the
Fresnel formula in the long--wavelength limit \be
\label{1}
R(\omega)=\left|\frac{\sqrt{\epsilon(\omega)} - 1}{\sqrt{\epsilon(\omega)}
 + 1}\right|^2
\ee
was applied where the frequency $\omega$ has to be taken at the laser 
frequency $\omega_{l}=1.8\,\cdot\,10^{15}$ Hz. 
The complex frequency--dependent dielectric permittivity
\be \label{def0}
\epsilon(\omega) = 
 1 +{i  \over \epsilon_0 \omega} \sigma(\omega) =
 1 - {\omega_{\rm pl}^2
  \over \omega [\omega + i \nu( \omega)]}
\ee
has been related to the dynamical conductivity $\sigma(\omega)$ or the 
dynamical collision frequency $\nu( \omega)$. 
The Drude formula 
\be \label{drude}
\sigma(\omega)\,=\,\frac{\epsilon_{0}\,\omega_{\rm pl}^{2}}
                   {\nu(0)-i\omega}
 \ee
follows 
if the collision frequency is taken in the static  
limit  
$\nu( 0)= \epsilon_0 \omega_{\rm pl}^2 / \sigma(0)$ 
relating it to the  static conductivity $\sigma(0)$. 
Different expressions for $\sigma(0)$ have been considered in
\cite{Mint}, but no satisfying explanation of the experimental
results has been given there.
As will be seen in the following discussion, the Drude model is not an
appropriate approximation under the experimental conditions
considered. In the present work, improvements in the calculation of
the dielectric function as well as considerations concerning the shape
of the wave front will be discussed to find a consistent theoretical
approach to the measured reflectivities.

\section{Reflection by a step-like plasma front}

In this section, we investigate the reflection coefficient at a shock
wave front where its width $d$ shall be neglected.  
According to the Fresnel formula, the reflection coefficient  
at a step-like plasma front
\cite{ref}
\be \label{fresnel}
  R(\omega)  =  \left| \frac{ 1-Z(\omega)}{1+Z(\omega)} \right|^2 \;\;\;,
\ee
 is related to 
the surface impedance $Z$ (normal incidence).
The surface impedance can be related to the transverse
dielectric function via
\be \label{imped}
\label{z}
  Z(\omega)  =  - \frac{i \omega}{\pi c} \int_{-\infty}^{\infty}
  \!dk\,\frac{1}{k^2-\omega^2 \epsilon_{t}(k, \omega) /c^2}\;\;.
\ee
This is a more
general expression than (\ref{1}) for the reflectivity. 
Assuming that the transverse dielectric function $ \epsilon_{t}(k,
\omega)$ is independent of the wave vector $k$, the integral 
(\ref{imped}) can be
evaluated leading to the expression (\ref{1}) given above.

In general, the dielectric function is related to a
dynamical, nonlocal collision frequency $\nu(k,\omega)$, 
\be
\epsilon_t(k,\omega) 
\,= \,1 - {\omega_{\rm pl}^2
  \over \omega [\omega + i \nu_t(k, \omega)]}\;,
\ee
which is defined by extending the above definition (\ref{def0}) 
to finite values of the wave vector $k$.
 In recent
papers \cite {RRRW,SWRRPZ}, an approach to the dielectric
function within a linear response theory was developed. 
 Different approximations  have been investigated with respect to 
 their consistency.
Having this in mind, we are able  to give a more
precise description of the dielectric function at the plasma
parameters considered here.

\subsection{Born approximation}

Before considering the nonlocal dielectric function below, we 
discuss the long--wavelength limit $k \to 0$, replacing 
 $\epsilon_t(k,\omega)$ by
$\epsilon_t(0,\omega)$ in Eq. (\ref{z}). In this case,  the
transverse and longitudinal dielectric function are identical. The
dynamical collision frequency can be evaluated in Born approximation
with respect to the statically screened potential (Debye potential),
see \cite{RRRW}, as (the non-degenerate case is considered)

\be \label{born}
  \nu^{\rm Born}(\omega) =-i g\,n_{e}\,  \int_0^\infty dy
  {y^4 \over (\bar n +y^2)^2} \int_{-\infty}^\infty dx e^{-(x-y)^2} {1
    - e^{-4 xy} \over xy (xy-\bar \omega -i \eta)}\,\,, 
\ee
where
\bea \nonumber
\bar n &= &{\hbar^2 n_e e^2 \over 8 \epsilon_0 m_e (k_BT)^2} \,\,,
\\ \nonumber
g &= & {e^4 \beta^{3/2} \over 24 \sqrt{2} \pi^{5/2} \epsilon_0^2
  m_e^{1/2}} \,\,,
\eea
and $\bar \omega = \hbar \omega / (4 k_BT)$.
The second integral in (\ref{born}) is a complex quantity,
\bea \nonumber
 \int_{-\infty}^\infty dx e^{-(x-y)^2} {\cal P} {1
    - e^{-4 xy} \over xy (xy-\bar \omega)} \,\,+ 
    \,\,i \,\,\pi  e^{-(\bar \omega/y-y)^2} {1 
    - e^{-4 \bar \omega} \over y\,\bar \omega}\,\,.
\eea

Evaluating the collision frequency
in the static 
limit ($\omega=0$), the result is the Ziman formula \cite{ziman} applied to a 
Debye potential in the case of  non--degeneracy,
\be \label{ziman}
\nu^{\rm Born}(0)\,=\,4\pi\,g\,n_{e}\, 
       \int_0^\infty dy {y^3 \over (\bar n +y^2)^2} e^{-y^2}\;\;.
\ee
In the following Tab. \ref{tab2}, the resulting values $R^{\rm Born}_{\rm dc}$ 
for the reflectivity 
calculated from the Drude formula (\ref{drude}) with the static 
collision frequency (\ref{ziman}) are
 compared with the reflectivities $R^{\rm Born} $ resulting from
the dynamical collision frequency in Born
approximation (\ref{born}) taken at the laser frequency $\omega_{l}$.

\begin{table}[b]
\begin{center} \begin{tabular}{|r|c|l|l|c|}
\hline
$P$,GPa & $ \omega_{\rm
  pl}/\omega_{l}$ &   $R^{\rm Born}_{\rm dc}$ & $R^{\rm Born} $
   & $R^{\rm Mermin, Born}_{\rm dc}$ 
\\  \hline
1.6  & 1.33 &0.272 & 0.304  & 0.272\\
3.1  & 1.77 &0.342 & 0.351  & 0.351\\
5.1  & 2.10 &0.381 & 0.380  & 0.376\\
7.3  & 2.36 &0.404 & 0.399  & 0.409\\
10.5 & 2.64 &0.429 & 0.419  & 0.443\\
16.7 & 2.99 &0.457 & 0.447  & 0.478\\ 
\hline
        \end{tabular} \end{center}
        \caption{Reflectivities from step-like density profiles,
          calculated at the experimental parameter values. $R^{\rm
            Born}_{\rm dc}$ -  Drude formula (\ref{drude}) with static 
collision frequency (\ref{ziman}), $R^{\rm Born} $ - dynamical
collision frequency in Born 
approximation (\ref{born}), $R^{\rm Mermin, Born}_{\rm dc}$ -
Eq. (\ref{z}) with the Mermin nonlocal 
dielectric function and static collision frequency in Born 
approximation. }
        \label{tab2}
\end{table}

 Compared with the observed  reflectivities, the results obtained in
Born approximation are too small. Similar values were also reported in 
\cite{Mint}.
As well known, the Born approximation (Faber-Ziman result)
underestimates the value of the dc conductivity. The correct
low-density value of $\sigma_{\rm dc}$ is given by the Spitzer formula
and can be obtained using a renormalization factor as discussed in
\cite{RRRW}, see also discussion in Sec.\,2.3 below.
 The use of the dynamical conductivity increases the
reflectivity by about 15 \%, but it also fails to produce the
steep dependence of the reflectivity on the electron density as
observed in the experiment, see Tab.\,\ref{tab1}.

\subsection{Nonlocal conductivity}

The general expression (\ref{fresnel}), (\ref{z}) for the reflection
coefficient contains an integral over the $k$ dependent transverse dielectric
function. We will discuss the effect of the nonlocal conductivity
using the Mermin approximation as given in \cite{SWRRPZ,andres}. The 
RPA solution is extended by introducing a complex frequency argument 
which contains the collision frequency. The Mermin expression for the 
dielectric function obeys particle number conservation. For the 
transverse dielectric function we obtain \cite{andres}
\bea 
\epsilon_{t}^{\rm Mermin}(k,\omega)&=&
         \epsilon_{l}^{\rm Mermin}(k,\omega) - \left( 
         \frac{c\,k}{\omega}\right)^{2}\,\left(1-{1 \over 
         \mu^{\rm Mermin}(k,\omega)}\right)\,\,,
\eea
with
\bea
\epsilon_{l}^{\rm Mermin}(k,\omega)&=&1+\frac{
   \left(1+i{\nu(\omega) \over \omega}\right)\,
    \left[\epsilon_{l}^{RPA}(k,\omega+i\nu(\omega))-1\right]
        }    {
    1+i{\nu(\omega) \over \omega}
    \left[ \epsilon_{l}^{\rm RPA}(k,\omega+i\nu(\omega))-1\right]/
       \left[\epsilon_{l}^{\rm RPA}(k,0)-1\right]
     }\,\,, \nonumber \\ \nonumber  \\ \nonumber
\mu^{\rm Mermin}(k,\omega)&=&1+\frac{
   \left(1+i{\nu(\omega) \over \omega}\right)\,
    \left[\mu^{\rm RPA}(k,\omega+i\nu(\omega))-1\right]
        }    {
    1+i{\nu(\omega) \over \omega}
    \left[ \mu^{\rm RPA}(k,\omega+i\nu(\omega))-1\right]/
       \left[\mu^{\rm RPA}(k,0)-1\right]
     }\,\,, \\ \nonumber  \\ \nonumber
  \mu^{\rm RPA}(k,\omega)&=&\frac{ 1}{1- \left( \frac{\omega}{c\,k} \right)^2  
           \left( \epsilon_{l}^{\rm RPA}(k,\omega)-
          \epsilon_{t}^{\rm RPA}(k,\omega)\right) }
\eea
The RPA expressions for the dielectric function of a Maxwellian plasma 
are
\bea
 \epsilon_{l}^{\rm RPA}(k,\omega)&=&1\,+\,{\kappa^{2} \over k^{2}}\,
       (2+ z_{e}D(z_{e})+z_{i}D(z_{i}))\;,\\ \nonumber
\epsilon_{t}^{\rm RPA}(k,\omega)&=&1\,+\,{1 \over \omega^{2}}\,
       (\omega_{{\rm pl},e}^{2}D(z_{e})+\omega_{{\rm pl},i}^{2}D(z_{i}))\;,
\eea   
with 
\be
z_c  = {\omega \over k}\, \sqrt{m_c \over 2 k_B T}\;,\qquad\quad 
\omega_{{\rm pl},c} ^{2}= {n_{c} e^{2}\over \epsilon_{0}m_{c}}\,,
\qquad c=e,i
\ee
and the Dawson integral
\be
D(z) = i \pi^{1/2} e^{-z^2} [1 + {\rm Erf}(iz)]\,\,.
\ee 

Using the static collision frequency (\ref{ziman}), the nonlocal
dielectric function was calculated.  The results $R^{\rm Mermin,
  Born}_{\rm dc}$ are also given in Tab.\,2.  There is no
essential modification if the $k$-dependence of the dielectric
function is taken into account. This can be explained by the fact
that, at the conditions given, the main contributions to the integral
over $k$ come from the region of $k \approx 0.0001 a_B^{-1}$. 
A comparison of 
the mean free path and the much larger skin depth also shows, that 
nonlocal effects are not relevant here. We conclude that
neither the account of the dynamical properties of the collision frequency
nor the account of the $k$ dependence in the dielectric function will
essentially modify the calculated reflectivity.  In particular, the
discrepancy between the calculated and measured reflectivity as a
function of the electron density, see Tab.\,\ref{tab1}, cannot be cured.

\subsection{Strong collisions and renormalization}

The Born approximation can and should be improved considering strong collisions
and renormalization as described, e.g., in \cite{RRRW}. 
Restricting ourselves to the static case, we discuss different 
approximations for $\sigma_{\rm dc}$. We
introduce dimensionless parameters $\Gamma, \Theta$ characterizing the
non--ideality and degeneracy, respectively,  of the plasma:
\bea \label{pp}
\Gamma &=& {e^2 \over {4\pi\epsilon_0 k_BT}}
 \left({4\pi n_{e} \over 3}\right)^{\!1/3} \,, \qquad\qquad
 \Theta \,=\,{2m_ek_BT \over \hbar^2}(3 \pi^2 n_{e})^{-2/3}\,.\nonumber
\eea
The dc conductivity for the statically
screened Coulomb potential  in Born approximation $\sigma_{dc}^{\rm Born}$
was already given in Sec.\,2.1 via (\ref{ziman}). In the  low-density 
limit, the dc-conductivity reads
\bea
\sigma_{\rm dc}^{\rm Born}&=& 0.299 {(4\pi \epsilon_0)^2(k_B
  T)^{3/2} \over {e^2
   m^{1/2}}} \left[{1\over 2}\, \ln {\Theta \over \Gamma} \right]^{-1} \,.
   \label{bh}
\eea
However, the correct expression for the dc conductivity in this
 limit is given by the Spitzer formula
\bea
\sigma_{\rm dc}^{\rm Spitzer}&=& 0.591 {(4\pi \epsilon_0)^2(k_B
  T)^{3/2} \over {e^2
   m^{1/2}}} \left[- {3 \over 2}\,\ln \Gamma \right]^{-1} \,.
   \label{spitzer}
\eea
It is obtained by taking into account strong collisions which modify
the Coulomb logarithm in (\ref{bh}). The prefactor also changes from 
taking into account higher moments
of the distribution function.

\begin{table}[b]
\begin{center}
\begin{tabular}{|r|r|r|r|r|r|r|r|}
\hline
$P$/GPa &  $\sigma_{\rm exp}$ &
           $\sigma^{\rm Born}_{\rm dc}$ &  
           $\sigma^{\rm ERR}_{\rm dc}$ &  
           $R^{\rm   ERR}_{\rm dc}$   
       \\ \hline
1.6  &  72 000 &  45 300.  &  90 000 & 0.502  \\
3.1  &  82 000 &  59 400.  & 125 000 & 0.586 \\
5.1  &  97 000 &  72 400.  & 160 000 & 0.629 \\
7.3  &  97 000 &  83 100.  & 195 000 & 0.660 \\
10.5 &  97 000 &  95 500.  & 240 000 & 0.691\\
16.7 & 100 000 & 114 000.  & 311 000 & 0.728 \\
\hline
\end{tabular}\end{center}
\caption{ Values of the dc conductivity in $(\Omega {\rm m})^{-1}$ 
for different approximations.
 $\sigma_{\rm exp}$  --  experimental estimates \cite{priv}, 
$\sigma^{\rm Born}_{\rm dc}$ -- static Born approximation 
(\ref{ziman}),
$\sigma^{\rm ERR}_{\rm dc}$ -- interpolation formula (\ref{interpol}) 
and corresponding reflectivity $R^{\rm   ERR}_{\rm dc}$
calculated using Drude formula (\ref{drude}).}
\label{tab3}
\end{table}

Recently, an interpolation formula for the dc conductivity of a fully ionized
Coulomb plasma was derived \cite{esser},
\bea\ \label{interpol} 
\sigma_{\rm dc}^{\rm ERR} &=& a_0 T^{3/2}\left(1+{b_1 \over \Theta^{3/2}}\right) 
 \left[
 D\ln(1+A+B)-C-{b_2 \over b_2+\Gamma\Theta}
 \right]^{-1} 
\eea
where $T$ in K, $\sigma$ in ($\Omega$m)$^{-1}$, and with the functions
\bea
A &=& \Gamma^{-3}{1+a_4/\Gamma^2 \Theta \over 1+a_2/\Gamma^2
 \Theta+a_3/\Gamma^4 \Theta^2} \;
 \left[a_1+c_1 \ln (c_2 \Gamma^{3/2}+1) \right]^2
 \,,\nonumber\\
B &=& b_3(1+c_3 \Theta) /\Gamma \Theta/(1+c_3 \Theta^{4/5})
 \,,\nonumber\\
C &=& c_4 / (\ln(1+\Gamma^{-1})+c_5 \Gamma^2 \Theta)
 \,,\nonumber\\
D &=& (\Gamma^{3}+a_5(1+a_6 \Gamma^{3/2})) /(\Gamma^{3}+a_5)
 \,.\nonumber
\eea

The  set of parameters is given by
$a_0=0.03064$, $a_1=1.1590$, $a_2=0.698$, $a_3=0.4876$, $a_4=0.1748$,
$a_5=0.1$, $a_6=0.258$, $b_1=1.95$, $b_2=2.88$, $b_3=3.6$, $c_1=1.5$,
$c_2=6.2$, $c_3=0.3$, $c_4=0.35$, $c_5=0.1$.
They are fixed by  the
low-density expansion of the dc conductivity (\ref{spitzer}), the 
strong degenerate limit and numerical data in 
for the dc conductivity the intermediate parameter region.

Using the Drude formula (\ref{drude}) with the  conductivities 
$\sigma^{\rm ERR}_{\rm dc}$ we obtain 
the reflectivities $R^{\rm ERR}_{\rm dc}$, see Tab.\,\ref{tab3}.
For comparison, experimental values for the dc conductivity 
\cite{priv} are given. The fit formula seems to overestimate the 
conductivity. The results for the reflectivities  $R^{\rm RR}_{\rm 
dc}$ are rather high, even exceeding the measured values. Obviously, 
the size of the calculated reflectivities can be shifted considerably 
according to the approximations made for the dc conductivity. But the 
strong increase of the measured values can not be explained yet.

It is possible to extend the approach given here for the dc
conductivity to the dynamic conductivity, as shown in \cite{RRRW}
using the Gould-DeWitt approach. However, as already shown for the Born
approximation, which is part of the Gould-DeWitt approach, we do not
expect a significant modification of the density dependence of the
reflection coefficient on the electron density. We also do not expect
a significant modification if the nonlocal, $k$ dependent dielectric
function is considered, because only the long-wavelength limit
contributes in calculating the impedance. Furthermore, within a more 
accurate treatment the contribution of the interaction with neutral 
atoms should be included as well. 
However, at the temperatures considered here the influence of 
neutrals on the conductivity is small.

In conclusion of this Section, we could show that improvements in 
 the theory of the dielectric
function lead to  substantial modifications in the reflectivity. 
For the parameter values considered here, the theory predicts
a high value of the reflectivity close to the values obtained from the
interpolation formula (\ref{interpol}). However, we are still 
not able to describe the strong
variation of the measured reflectivity with the electron density.

\section{Smooth density profile}

Obviously, it is not possible to interpret the measured values of the
reflection coefficient of dense Xenon plasmas within the assumption of 
a step-like density profile, i.e. $d \approx 0$ for width of the shock wave
front. In particular, the steep increase of the reflectivity only at 
densities above the critical one can not be explained despite a highly 
sophisticated approach to the calculation of the dielectric function.

Using the interpolation formula for the dc--conductivity 
(\ref{interpol}) and the 
Drude formula (\ref{drude}),
 effective densities necessary to reproduce
the measured values for the reflection coefficient can be deduced. 
They have been found to lie between 0.75 and 1.6 of the critical 
density $n_{\rm cr} =1.02 \times 10^{21}$ cm$^{-3}$ where the plasma frequency
coincides with the frequency of the probe laser.

 These effective densities  can be
considered as an argument that the reflection of electromagnetic
radiation occurs already in the outer region where the density is
low. Within a simplified picture, considering a profile with
increasing electron density, the radiation penetrates the low-density
region of the plasma up to the region where the density  approaches
the critical value. Here the wave will be reflected.

To perform an exploratory calculation, a density profile was
assumed where the electron density increases linearly with distance $z$ from 
zero up to the saturation value $n_e$ at the distance $d$. We
assume the following linear dependence of the dielectric function on 
the distance of the shock wave front $z$ 
\begin{equation}
        \epsilon(z)\,=\,1\,-\,\frac{z\, \omega_{\rm pl}^{2}}
        {L \,\omega^{2}\left(1+{i\nu_{\rm cr} \over \omega}
\right)} \,\,,
\end{equation}
$L$ is the depth where the critical density 
$n_{\rm cr}$ is 
reached and the radiation is assumed to be reflected. The 
total width of the shock wave front is than determined by
\begin{equation}
        d=\frac{n_{e}}{n_{\rm cr}} \,L\,\,.
      \label{breite}
\end{equation}
The collision frequency $\nu_{\rm cr} = 4.09 \times 10^{14}$ s$^{-1}$ was 
determined from the static collision frequency at the critical density
using the interpolation formula for the static conductivity (\ref{interpol}).
The reflectivity can be calculated via ( see \cite{kruer})
\begin{equation}
        R=\exp\left(-\frac{8 \nu_{\rm cr} L}{3 c}\right)\,.
       \label{rcr}
\end{equation}
Taking the experimental values $R$ for the reflectivity, 
the width $d_{\rm cr}$ of the shock wave front according to (\ref{rcr}) 
is given in  Tab.\,\ref{tab4}.

\begin{table}
\begin{center} \begin{tabular}{|r|l|c|c|c|c|l|}
\hline
$P$,GPa &  $R$ & $L/(10^{-7}{\rm m})$ & $d
  /(10^{-6}{\rm m})$ & $d_{\rm lin}/(10^{-6}{\rm m})$\\
\hline
1.6  &  0.096  & 6.01 & 1.13 & 1.26\\
3.1 & 0.12  & 5.44 & 1.82 & 2.08\\
5.1  & 0.18  & 4.40 & 2.07 & 2.54\\
7.3  & 0.26  & 3.46 & 2.06 & 2.68\\
10.5 & 0.36  & 2.62 &  1.95 & 2.56\\
16.7 & 0.47  & 1.94 & 1.84 & 2.12\\
\hline
\end{tabular} \end{center}
\caption{ Calculation of the width $d$ of the shock wave front from
  the reflectivity $R$, 
assuming a linear density profile. The critical 
density $n_{\rm cr}$ is reached at depth $L_{\rm cr}$. $d_{\rm
  cr}$ according to Eq.\,(\ref{breite}), 
  $d_{\rm lin}$ solving the propagation of 
radiation in a linear density profile.}
\label{tab4}
\end{table}
 
Compared with the value given in \cite{Mint}, see also Sec.\,1 above, 
our estimation of the width of the shock wave front is larger by about 
one order of magnitude and almost independent of the thermodynamic 
parameters. Only the value at the lowest density value comes 
out to be smaller.

A more rigorous treatment of the propagation of laser radiation through 
a shock wave front should take into account the dependence of the collision 
frequency on the local electron density. An arbitrary density profile 
$n_e(z)$ can be approximated by a sufficiently large number of thin layers 
with constant electron density, and solving the boundary 
conditions when going from one slab to the 
next one. In such a way, any arbitrary density profile $n_e(z)$ can be 
calculated.

We have considered a linear dependence of the electron density on the 
distance $z$ within the shock wave front of width $d$, $n_e(z)= n_e\,z/d$, 
with $n_e$ being the electron density behind the shock front. 
The subdivision of the width $d$ into equidistant slabs has been increased 
until convergence was reached. For given density $n_e$, the 
interpolation formula for the dc conductivity $\sigma_{\rm dc}^{\rm ERR}$ 
has been used to find the dielectric function of the corresponding slab 
according to the Drude formula (\ref{drude}). 
Results $d_{\rm lin}$ reproducing the 
experimental reflectivity values $R$ are shown in Tab.\,\ref{tab4}. 
Convergence was reached by dividing $d_{\rm lin}$ into 16 equidistant slabs.

\section{Conclusion}

In order to infer plasma parameters from optical reflection
coefficient measurements of dense Xenon plasmas, we propose a width of the
shock wave front of about $2\cdot10^{-6}$ m. Only at the lowest density
measured, the value for $d$ may be smaller. The linear density 
profile should be considered as a model only to estimate the width of the shock 
wave front. Our method to approximate a given density profile by a sufficiently 
large number of thin layers with constant electron density can be applied
to any $n_e(z)$ which, in general, could be obtained by
determining the distribution function for the 
non-equilibrium process of the shock wave propagation.

Our calculations are based on an 
interpolation formula for the dc conductivity, obtained from a systematic 
quantum statistical treatment of limiting cases. In particular, the account 
of the renormalization factor and of strong collisions is essential to obtain 
the correct low-density limit. The uncertainty in using the interpolation 
formula increases for $\Gamma \le 1$. The values for the plasma 
parameter $\Gamma$ for the dense Xenon plasmas considered here are in the 
region $ 1 < \Gamma < 2$, and $5 > \Theta >1.5$ and we estimate the error of 
about 30 \%. Furthermore, the dynamical 
collision frequency should be used instead of the static one. Using the 
Gould-DeWitt approach, the dynamical collision frequency was investigated in 
\cite{RRRW}, and minor modifications (about one half of that given above) are 
expected. The  effects of non-locality can be 
neglected in the region of plasma parameters considered, as shown above for 
the Born approximation.

Xenon under the conditions considered is a partially ionized
plasma. The composition is shown in Tab.\,\ref{tab1}. 
The conductivity as well as 
the related quantities are influenced by the neutral component, which
leads to a modification of the reflection coefficient. However, at the 
temperatures of about 30 000 K the contribution of the neutral
component to the conductivity is small and will not be considered
here. Similarly, the role of non--equilibrium effects such as relaxation
of the composition will also not be considered here. 

 Shock wave front
investigations may be the subject of forthcoming experimental work.
An interesting point would be the simultaneous determination of reflectivities
at different frequencies. It is expected that in this way more information 
about 
the density profile of the shock wave front can be obtained.

 \section*{Acknowledgments}
We would like to thank the DFG for support within the
SPP 1053 ``Wechselwirkung intensiver Laserfelder
mit Materie" and the SFB 198 ``Kinetik partiell ionisierter Plasmen" as 
well as via the  
Grant RFBR-DFG No 99-02-04018. H.R. acknowledges a DFG research 
fellowship.

 \begin{received}
 Received 
  \today. 
 \end{received}

\end{document}